\documentclass{article}
\usepackage{amsmath}
\usepackage{amssymb}
\usepackage{amsthm}
\usepackage{graphicx}
\usepackage{psfrag}
\usepackage[hang, nooneline]{subfigure}
\usepackage{color}

\usepackage{soul} 
\usepackage{cite}

\newcommand{\BS} {\mathbf{S}}
\newcommand{\mg}{ \mathfrak{g}}
\author{Natascha Riahi\footnote{e-mail address: natascha.riahi@gmx.at}
\\ University of Vienna, Faculty of Physics, Gravitational Physics
\\ Boltzmanng. 5, 
1090 Vienna, Austria \vspace*{1ex} \\
Monika E. Pietrzyk\footnote{e-mail address: m.pietrzyk@exeter.ac.uk }
\\ Mathematics and Physical Sciences, University of Exeter, 
 EX4 4QL Exeter, UK
\\ 
 }

\date{}
\title{On the relation between the canonical Hamilton-Jacobi equation  
and the De Donder-Weyl  Hamilton-Jacobi formulation in general relativity }

\begin{document}
\maketitle
\begin{abstract}
We discuss the relation between the canonical Hamilton-Jacobi theory and the 
De Donder-Weyl Hamilton-Jacobi theory known in the calculus of variations using the 
examples of a scalar field in curved space-time and general relativity. 
By restricting ourselves to the Gaussian coordinates we show how the canonical 
Hamilton-Jacobi equation of general relativity can be derived from the 
De Donder-Weyl Hamilton-Jacobi formulation of the Einstein equations.  

\end{abstract}

\section{Introduction}
The De Donder-Weyl (DW) theory \cite{Weyl, DD} is a generalization of the Hamiltonian formulation to field theory which  does not distinguish between space and time. Given a
Lagrangian $L(y^{a}, y^{a}_{\mu},x^{\nu})$ depending on
 the space-time variables $x^{\nu}$, 
  field variables $ y^{a}$, and their first jet coordinates $y^{a}_{\mu}$ (such that a restriction to
  a field configuration $y^{a}=y^{a}(x^{\nu})$ implies: $y^{a}_{\mu}= \partial y^a (x) /\partial x^{\mu} =:  \partial_{\mu} y^{a}$), a Legendre transformation to  new variables 
   $ p^{\mu}_{a}:=
   \frac{\partial L}{\partial y^{a}_{\mu}}$    
(polymomenta) 
and $H :=y^{a}_{\mu}\, p^{\mu}_{a}-L $ (the DW Hamiltonian) 
 enables us to write the Euler-Lagrange field equations in the DW Hamiltonian form 
\begin{equation}
\label{DWH}
 \partial_{\mu}y^{a}=\frac{\partial H}{ \partial p^{\mu}_{a}}\,,
 \quad
 \partial_{\mu}p^{\mu}_{a}=
 -\frac{\partial H}{ \partial y^{a}}\,,
\end{equation}
provided the regularity condition 
$\det\left(\frac{\partial^2 L}{\partial y^a_\mu \partial y^b_{\nu} }\right) 
\neq 0$ 
is fulfilled.
The  dynamical content of the DW Hamiltonian formulation 
 can be represented also by the  DW Hamilton-Jacobi (DWHJ) equation \cite{Weyl,DD,Rieth}
\begin{equation}
 \label{DWHJ}
  \partial_{\mu}S^{\mu}+H\left(y^{a}, \frac{\partial S^{\mu}}{\partial y^{a}},x^{\mu}
  \right)=0\, .
\end{equation}
This partial differential equation for the eikonal functions $S^{\mu}(y^{a},x^{\mu})$\,
determines 
 the solutions of (\ref{DWH}) by the  embedding conditions
\begin{equation}
 \label{conditions}
 \frac{\partial L}{\partial y^{a}_{\mu}} = 
 \frac{\partial S^{\mu}}{\partial y^{a}} .
\end{equation}
Geometrical aspects of DWHJ equation have been studied recently in \cite{Campos,deleon-hj}. 
The historical role of the HJ formulation of mechanics in the discovery of the 
Schr\"odinger equation \cite{Schr} makes the DWHJ formulation particularly interesting. 
In fact, within the framework of precanonical quantization which uses the DW theory instead of the 
canonical Hamiltonian formalism \cite{igor-pre,PrecF}  it was already shown that the 
DWHJ equation for scalar fields follows from the corresponding precanonical 
generalization of the Schr\"odinger equation 
in the classical limit \cite{PrecCL}. 
Precanonical quantization has been applied to quantum Yang-Mills theory \cite{igor-ym2,igor-ym3},
quantum scalar field theory in curved spacetime \cite{icurv1,icurv2,icurv3}, 
and to quantization of gravity in metric \cite{PrecGR1,PrecGR2} and vielbein variables \cite{PrecGRN,qg4}. 
 
 To understand the connection between this approach and the canonical quantization 
of general relativity \cite{Kiefer,Rovelli}, we investigate here a relationship 
between the DWHJ equation for general relativity \cite{DD,Horava} and the canonical HJ equation
\cite{Peres} that has been used to explore the semiclassical approximation of canonical quantum gravity \cite{KS,LM,Reginatto}. In section 2, we establish a relation between the DWHJ equation 
and the canonical HJ equation for a scalar field in a general curved space-time. In section 3, 
in Gaussian coordinates, we derive the canonical HJ equation for general relativity from the DWHJ 
formulation.  Our results generalize the relation between the DWHJ and canonical HJ equation 
in flat space-time found by Kanatchikov \cite{ScalarFieldM} 
and applied to the bosonic string by Nikoli\'c \cite{nikolic}.

 

\section{Canonical HJ vs. DWHJ for the scalar field in curved space-time} 

The Lagrangian density of a scalar field in curved spacetime with the metric $g_{\mu\nu}$
$(g:=\det g_{\mu \nu})$
 reads 
\begin{equation} \label{la}
L =
- \frac{1}{2} 
g^{\mu\nu} \varphi_\mu \varphi_\nu\sqrt{-g} - V(\varphi)  \sqrt{-g} 
 \,.
\end{equation}
The canonical HJ equation is derived by using 
a foliation of spacetime 
by 
space-like hypersurfaces~$\mathcal{F}$ labelled by the time function $t$.
In adapted coordinates, 
the metric $g_{\mu\nu}$ 
decomposes in terms of the lapse function $N$, shift vector $N^{i}$ and the spatial components $h_{ij}$ of the first fundamental form $h_{\mu \nu}$
\begin{equation}
\label{3+1}
ds^2= \left(N_{i}N^i-N^2\right)dt^2+2 N_{i}dx^{i}dt
+h_{ij}dx^idx^j\,.
\end{equation}
By introducing the canonical momentum $\pi$ and the canonical Hamiltonian density $\mathcal{H} (\phi(\mathbf{x}),\pi(\mathbf{x}))$ in the usual way (cf. e.g. \cite{Long}),
the canonical HJ equation for the eikonal functional $\BS(\varphi(\mathbf{x}),\mathbf{x}, t)$
takes the explicit form  ($h = \det \, h_{i j}$)
\begin{equation}
\label{SHJ}
 \partial_t 
 \BS +\int\! d\mathbf{x} \,
 \left(\frac{N}{2 \sqrt{h}}\, \frac{\delta \BS}{\delta \varphi(\mathbf{x})} \frac{\delta \BS}{\delta \varphi(\mathbf{x})}
 + \frac{1}{2} 
 h^{i j}\partial_i\varphi \partial_j\varphi \sqrt{-g}   +
 V(\varphi)  \sqrt{-g} 
 +N^{i}\partial_i \varphi \frac{\delta \BS}{\delta \varphi(\mathbf{x})}\right)=
0\, ,
\end{equation} 
where the solutions of field equations are embedded 
by the condition 
$ \sqrt{-g}  g^{0\mu} \partial_\mu \varphi =  - \frac{\delta \BS}{\delta \varphi(\mathbf{x})}$.

On the other hand, from (\ref{la}) we obtain the polymomenta  and the DW Hamiltonian density 
\begin{equation}
 \label{ScalarPoly}
 p^{\mu}= \frac{\partial L}{\partial \varphi_{\mu}}= -\sqrt{-g}\,\varphi_{\nu}\, g^{\mu \nu}\,,
 \quad H=-\frac{1}{2 \sqrt{-g}}\, p^{\mu}p^{\nu}g_{\mu \nu}
 +  V(\varphi) \sqrt{-g}\,\,, 
\end{equation}
 and the DWHJ equation (\ref{DWHJ}) 
 for the eikonal densities $S^{\mu}(\varphi, x^{\nu})$ 
\begin{equation} 
 \label{SDWHJ}
 \partial_{\mu}S^{\mu}-\frac{1}{2 \sqrt{-g}}\, \frac{\partial S^{\mu}}{\partial \varphi}
 \frac{\partial S^{\nu}}{\partial \varphi}g_{\mu \nu}
 + V(\varphi)\sqrt{-g} = 0 \, , 
\end{equation}
where $p^\mu = \frac{\partial S^\mu}{\partial \phi}$ and 
the solutions of classical field equations are given by the embedding condition 
\begin{equation} \label{embed}
 \frac{\partial S^{\mu}}{\partial \varphi}=-\sqrt{-g}g^{\mu \nu}\partial_{\nu} \varphi\,.
\end{equation}

Our task is to understand the relationship between the canonical 
formulation (\ref{SHJ}), which requires a space-time split, 
and the DW formulation (\ref{SDWHJ}) where all space-time variables are treated 
equally. Here we generalize the consideration of \cite{ScalarFieldM} 
to curved spacetime.   
At first we introduce the restriction of the densities 
 $S^{\mu}(\varphi, x^{\nu})$
to a field configuration $\varphi(\mathbf{x})$ on a hypersurface $\mathcal{F}$ at a time $t$, 
${ S^{\mu}|_{\varphi(\mathcal{F})} := S^{\mu}(\varphi(\mathbf{x}), \mathbf{x},t)}$.
The embedding condition (\ref{embed}) in adapted coordinates yields 
\begin{align}
 \label{Srelations}
 \frac{\partial S^{i}}{\partial \varphi}\Big{|}_{\varphi(\mathcal{F})}
 +N^{i}\frac{\partial S^{0}}{\partial \varphi}\Big{|}_{\varphi(\mathcal{F})}
 &=-\sqrt{-g}\, \partial_j\varphi\,  h^{i j}\, , 
\\ \label{Srelation0}
\frac{\partial S^{0}}{\partial \varphi}\Big{|}_{\varphi(\mathcal{F})}N^2 
&=\sqrt{-g}\left(\partial_{0}\varphi-
N^{i}\partial_{i}\varphi\right)\,.
\end{align}
Following \cite{ScalarFieldM} we construct the eikonal functional 
from the DW eikonal densities 
 \begin{equation}
 \label{Projektion}
 \BS:=\int \limits_{\mathcal{F}} \! d\mathbf{x}\, S^{0}|_{\varphi(\mathcal{F})}\,.
 \end{equation}
Then, using (\ref{SDWHJ}) 
and (\ref{Srelations}),  for the time derivative of $\BS$  we obtain
\begin{align}
\begin{split} \label{chj}
 \partial_t \BS=
\int \! d\mathbf{x}\, \partial_{t} S^{0}(\varphi(\mathbf{x}), 
\mathbf{x},t) =\int\! d\mathbf{x}
 \bigg\{
 -\frac{d}{dx^i}S^{i}\Big{|}_{\varphi(\mathcal{F})}
- \frac{N}{2 \sqrt{h}}\, \left(\frac{\partial S^{0}}{\partial\varphi}\Big{|}_{\varphi(\mathcal{F})} \right)^2 
\! & \\
-\left.
N^{i} \partial_{i}\varphi \frac{\partial S^{0}}{\partial\varphi}\Big{|}_{\varphi(\mathcal{F})}
 -
 \,\frac{1}{2} \sqrt{-g} h^{i j}\partial_{i}\varphi\partial_{j} \varphi -\sqrt{-g}\, V(\varphi (\mathbf{x}))
 \right\} \; & \,, 
 \end{split}
\end{align}
where 
 the notation for the total divergence of the eikonal density on ${\varphi(\mathcal{F})}$ is introduced 
\begin{equation}
\label{derivative}
 \frac{d}{dx^i}S^{i}|_{\varphi(\mathcal{F})}=
 \frac{\partial S^{i}}{\partial\varphi}
 \Big{|}_{\varphi(\mathcal{F})}
\frac{\partial\varphi(\mathbf{x})}{\partial x^i}
+\frac{\partial S^{i}}{\partial x^i}
\Big{|}_{\varphi(\mathcal{F})}\,.
\end{equation}
By noticing that the functional derivative of the 
 functional (\ref{Projektion}) with respect to $\varphi(\mathbf{x})$ 
 reads 
\begin{equation}
\label{Funktionalableitung}
 \frac{\delta \BS}{\delta \varphi(\mathbf{x})}
 =\frac{\partial S^{0}}{\partial\varphi}
 \Big{|}_{\varphi(\mathcal{F})}\,,
\end{equation}
and assuming that $S^{i}\mathop{|}_{\varphi(\mathcal{F})}$ 
vanishes at the boundary of ${\varphi(\mathcal{F})}$, so that the 
integral of the total divergence in (\ref{chj}) vanishes, we conclude that 
the functional $\BS$ obeys the canonical HJ equation (\ref{SHJ}) 
as the consequence of the DWHJ equation (\ref{SDWHJ})
for the eikonal densities $S^\mu$.

\section{The DWHJ equation and the canonical HJ equation in general relativity}

In $3+1$ dimensions, the DWHJ equation for general relativity 
 found by De Donder \cite{DD} and Ho\v{r}ava \cite{Horava} reads 
\begin{equation}
\label{GDWHJ}
  \partial_{\mu}S^{\mu}
  + 
  \mg^{\alpha \gamma}
  \left(\frac{\partial S^{\delta}}{\partial \mg^{\alpha \beta}}
  \frac{\partial S^{\beta}}{\partial \mg^{\gamma 
  \delta}}
  -\frac{1}{3}
  \frac{\partial S^{\beta}}{\partial \mg^{\alpha \beta}}
  \frac{\partial S^{\delta}}{\partial \mg^{\gamma 
  \delta}}
  \right)=0  \ . 
 \end{equation}
It uses the  metric density components 
$\mg^{\alpha \beta}=\sqrt{-g}\,g^{\alpha \beta}$ as the field variables, 
so that $S^\mu = S^\mu (\mg^{\alpha\beta}, x^\nu)$.  

The polymomenta derived from the truncated Hilbert action without the surface term
 are expressed in terms of the Christoffel symbols: 
\begin{equation}
 \label{GPoly}
  Q^{\alpha}_{\beta \gamma}=
 \frac{1}{2}\left( \delta^{\alpha}_{\beta}
 \Gamma^{\mu}_{ \gamma \mu}+
 \delta^{\alpha}_{\gamma}
 \Gamma^{\mu}_{ \beta \mu}\right)-
\Gamma^{\alpha}_{\beta \gamma}\,.
  \end{equation}  
The solutions of the Einstein's field equations are constructed 
from the eikonal densities $S^\mu$ using the embedding condition 
\begin{equation}
\label{Grelations} 
Q^{\alpha}_{\beta \gamma}= 
\frac{\partial   S^{\alpha}}
{\partial \mg^{\beta \gamma}}
=
 \frac{1}{2}\left( \delta^{\alpha}_{\beta}
 \Gamma^{\mu}_{ \gamma \mu}+
 \delta^{\alpha}_{\gamma}
 \Gamma^{\mu}_{ \beta \mu}\right)-
\Gamma^{\alpha}_{\beta \gamma}\, 
  \end{equation}
and the well known expression of the Christoffel symbols in terms of the first 
derivatives of the metric.   

 
 In order to understand the relation between the DWHJ equation for general relativity
 (\ref{GDWHJ}) and the canonical HJ equation found by Peres \cite{Peres}
 \begin{equation}
\label{PeresHJ}
\int\! d\mathbf{x}
\left(\sqrt{-g}\,\,^{3}\!R+\frac{1}{\sqrt{-g}}
 \left(\frac{1}{2}g_{ij}g_{kl}-g_{ik}g_{il}\right)
 \frac{\delta \BS}{\delta g_{ij}}\frac{\delta \BS}{\delta g_{kl}}
 \right)=0\,,
\end{equation}
we have to perform a space-time decomposition in (\ref{GDWHJ}). 
In this paper, we confine ourselves to the simpler case of   
adapted (Gaussian) coordinates with $g_{0i}=N_{i}=0$ and $g_{00}=-1$.
 Following \cite{ScalarFieldM} 
we construct 
the canonical HJ functional $\BS ([g_{ij}(\mathbf{x})],t)$ 
from the eikonal densities $S^\mu ({\mg{}^{\alpha\beta}, x^\mu})$ 
\begin{equation}
 \label{Projektion2}
 \BS ([g_{ij}(\mathbf{x})],t) :=\!\int \limits_{\mathcal{F}} \!d\mathbf{x}\, S^{0}|_{\mg(\mathcal{F})}
 =\!\int \limits_{\mathcal{F}} \! d\mathbf{x}\, S^{0}(\mg^{\alpha \beta}(\mathbf{x}),\mathbf{x},t), 
 \end{equation} 
where 
$ S^{\mu}|_{\mg(\mathcal{F})}\equiv
 S^{\mu}(\mg^{\alpha \beta}(\mathbf{x}),\mathbf{x},t) $
denotes the restriction of the eikonal densities 
$ S^{\mu}\left(\mg^{\beta\gamma},x^{\mu}\right)$ to the spatial field configurations $\mg^{\alpha\beta} (\mathbf{x})$ 
on a hypersurface $\mathcal{F}$ at the time $t$. 
In Gaussian coordinates,
the embedding conditions (\ref{Grelations}) 
 give rise to the following relations 
\begin{subequations}
\label{Gaussian}
  \begin{align}
&  \frac{\partial S^{i}}{\partial \mg ^{0 j}}+
 \delta^{i}_{j}
 \frac{\partial S^{0}}{\partial \mg ^{k l}}g^{k l}
 - \frac{\partial S^{0}}{\partial \mg ^{j k}}g^{k i}
=0\,,
\quad 
 \frac{\partial S^{i}}{\partial \mg ^{00}}=0\,, \quad \quad \quad 
 \frac{\partial S^{0}}{\partial \mg ^{0 0}}+
  \frac{\partial S^{0}}{\partial \mg ^{ij}}g^{ij}
 =0\,, \label{Integral}
 \\
\label{spatial}
&\frac{\partial S^{i}}{\partial \mg ^{j k}}= \frac{1}{2}\left( \delta^{i}_{j}
 \Gamma^{l}_{k l}+
 \delta^{i}_{k}
 \Gamma^{l}_{j l}\right)-
\Gamma^{i}_{j k}\,, \quad\, \; 
\frac{\partial S^{0}}{\partial \mg ^{0 i}}=
\frac{1}{2} \Gamma^{j}_{ij}\,, \\
\label{temporal}
& \frac{\partial S^{0}}{\partial \mg ^{i j }}= - \Gamma^{0}_{ij}=-\frac{1}{2}
\partial_{0} g_{i j}\,.
\end{align}
\end{subequations} 
Using (\ref{GDWHJ}) and (\ref{Integral},\ref{spatial})
we obtain for the time derivative of  $\BS$   
\begin{align}
\begin{split}  \label{dtS} 
\partial_{t} \BS=\int\! d\mathbf{x} \,
 \left\{
 -\frac{d}{dx^i}S^{i}\Big{|}_{\mg(\mathcal{F})}+
 \frac{\partial S^i}{\partial \mg^{\alpha \beta}}\Big{|}_{\mg(\mathcal{F})}\,\partial_i \mg^{\alpha \beta}
 + \mg^{ij} \left ( \Gamma^{l}_{k l} \Gamma^{k}_{ij}  
 - \Gamma^{k}_{i l} \Gamma^{l}_{j k} \right)
 \right. &
 \\
\qquad \,  + 
\ \frac{1}{\sqrt{-g}}
\left(\frac{\partial S^{0}}{\partial \mg^{ij}}\Big{|}_{\mg(\mathcal{F})}\,\mg^{ij}\right)^2 -
\,
\frac{1}{\sqrt{-g}} \,
\frac{\partial S^{0}}{\partial \mg^{ij}}\Big{|}_{\mg (\mathcal{F})} \,\frac{\partial S^{0}}{\partial \mg^{kl}}\Big{|}_{\mg(\mathcal{F})}
\,\mg^{ik} \mg^{jl}
\bigg\}&\, ,
\end{split}
\end{align}
where the  total divergence is understood as in (\ref{derivative}). 
 Then, by noticing that 
\begin{equation}
\frac{\partial S^{0}}{\partial \mg^{ij}}
 \Big{|}_{\mg(\mathcal{F})} = 
\frac{\delta \BS}{ \ \delta \mg^{ij}}   
\end{equation} 
 and 
 using (\ref{Integral}), (\ref{spatial}), and the identity 
\begin{equation}
\label{ddg}
 \frac{\partial}{\partial g_{\alpha \beta}}=
 \sqrt{-g}
 \left(-g^{\alpha \mu} g^{\beta \nu}+
 \frac{1}{2}g^{\mu \nu}g^{\alpha \beta}
 \right) \frac{\partial}{\partial \mg ^{\mu \nu}}\,,
\end{equation}
we conclude that, under the assumption that the surface terms do not contribute,  
the right hand side of (\ref{dtS}) coincides with the Hamiltonian constraint 
in the canonical HJ form (\ref{PeresHJ}). 
Consequently, on the surface of the Hamiltonian constraint, $\partial_t \BS = 0$ 
and hence the timelessness of the 
canonical formalism of general relativity emerges from the DWHJ formulation as a consequence 
of the space-time splitting. 

\section{Conclusions}

We derived the canonical functional derivative HJ equation 
from the partial derivative DWHJ equation for the scalar field theory in curved space-time 
and general relativity in metric variables. 
In both cases, the Ansatz  proposed in  \cite{ScalarFieldM},   
 which relates the canonical HJ functional with the DW eikonal functions/densities, 
 holds true.
In general relativity, where we confined ourselves to the case of the 
Gaussian coordinates, we derived the standard Hamiltonian constraint in the HJ 
form from the DWHJ equation. We expect that a consideration in general coordinates will 
also reproduce the momentum constraint. The obtained results 
 should be helpful for the comparison of 
canonical quantum gravity \cite{Kiefer,Rovelli} and precanonical quantization of general relativity \cite{PrecGR1,PrecGR2,PrecGRN,qg4}, and for the study of the latter in the semiclassical 
approximation, where it should reproduce the DWHJ equation. They may also be helpful for 
understanding the origin of the problem of time in quantum gravity. We also expect that the DW and 
the DWHJ formulation of the Einstein's equations can be useful for their numerical integration 
using the polysymplectic integrator which preserves the fundamental structure of the DW Hamiltonian 
form of field equations (cf. \cite{mp}). Our result can be viewed as a classical counterpart of the 
study of the relations between the functional Schr\"odinger representation in quantum field theory 
(see e.g. \cite{Long}) and the precanonical quantization using the DW Hamiltonian theory, which has been undertaken in \cite{igor-ym2,igor-ym3,icurv1,icurv2,icurv3,ScalarFieldM,iatmp} 
 and whose extension to quantum gravity is unknown so far.



\end{document}